\documentclass[prx,%
 reprint,
 amsmath,amssymb,
 aps,prx
]{revtex4-2}
\usepackage[english]{babel}

\usepackage{amsmath, amssymb, amsthm}
\usepackage{graphicx}
\usepackage{hyperref}
\usepackage{xcolor}
\usepackage{booktabs}
\usepackage{dsfont}

\begin{document}

\title{Habitat fragmentation promotes spatial scale separation under resource competition}

\author{James Austin Orgeron} 
\author{Malbor Asllani} 
 \affiliation{Department of Mathematics, Florida State University,
1017 Academic Way, Tallahassee, FL 32306, United States of America}

\begin{abstract}
Habitat fragmentation, often driven by human activities, alters ecological landscapes by disrupting connectivity and reshaping species interactions. In such fragmented environments, habitats can be modeled as networks, where individuals disperse across interconnected patches. We consider an intraspecific competition model, where individuals compete for space while dispersing according to a nonlinear random walk, capturing the heterogeneity of the network. The interplay between asymmetric competition, dispersal dynamics, and spatial heterogeneity leads to nonuniform species distribution: individuals with stronger competitive traits accumulate in central (hub) habitat patches, while those with weaker traits are displaced toward the periphery. We provide analytical insights into this mechanism, supported by numerical simulations, demonstrating how competition and spatial structure jointly influence species segregation. In the large-network limit, this effect becomes extreme, with dominant individuals disappearing from peripheral patches and subordinate ones from central regions, establishing spatial segregation. This pattern may create favorable conditions for speciation, as physical separation can reinforce divergence within the population over time.
\end{abstract}

\date{\today}

\maketitle

\section{Introduction}
\label{Sec:Intro}

Habitat fragmentation—the division of continuous ecosystems into smaller, spatially isolated patches—is a pervasive consequence of anthropogenic land-use change, with profound implications for biodiversity and ecological stability. By increasing edge effects, restricting dispersal, and reducing available habitat, fragmentation alters community composition and raises extinction risk, especially for specialists and dispersal-limited species \cite{wilcox1985conservation, fahrig2003effects, haddad2015habitat}. These ecological disruptions are further compounded by long-term evolutionary consequences: limited gene flow between patches can promote spatial genetic structuring, foster local adaptation, and under divergent selective pressures, lead to reproductive isolation and ecological speciation \cite{rundle2005ecological}. As such, fragmentation not only accelerates biodiversity loss but also shapes macroevolutionary dynamics, underscoring the need for spatially explicit models that capture both ecological interactions and evolutionary processes \cite{fischer2007landscape}.

A well-documented example of how multiple ecological pressures interact in fragmented landscapes is the displacement of native red squirrels (\emph{Sciurus vulgaris}) by invasive gray squirrels (\emph{Sciurus carolinensis}) in the United Kingdom \cite{murray}. This process is driven by a combination of factors, including competitive exclusion by gray squirrels \cite{gurnell2004alien}, disease-mediated impacts—particularly the spread of squirrelpox virus, which is lethal to reds but asymptomatic in grays \cite{tompkins2003ecological, rushton2000modelling}—and habitat fragmentation that disproportionately affects the more habitat-sensitive red squirrel \cite{lurz1995ecology}. Red squirrels require large, continuous coniferous forests and are less adaptable to fragmented or mixed woodlands, whereas gray squirrels, as generalists, thrive in disturbed and patchy environments. As suitable habitat becomes increasingly subdivided, red squirrels are confined to isolated refugia, while gray squirrels dominate the surrounding matrix. These ecological conditions foster spatial segregation, reduce gene flow, and under persistent divergent pressures, may initiate processes of adaptive divergence or even speciation. Motivated by this, we develop a two-species spatial model incorporating nonlinear density-dependent movement, local carrying capacities, and interspecific exclusion to study the emergence of stable segregation, coexistence, or exclusion in fragmented ecological systems.

While many studies on habitat fragmentation combine local demographic dynamics with dispersal, in this paper our focus is specifically on how spatial heterogeneity and carrying capacity constraints alone can drive competitive segregation. In this regime, population numbers are assumed to be locally regulated, and the primary dynamics of interest arise from movement and exclusion rather than growth or decline. This perspective aligns with movement ecology frameworks and nonlinear dispersal models that emphasize spatial redistribution over demographic processes \cite{shigesada1979spatial, shigesada1997biological, turchin1998quantitative, okubo2002diffusion, ovaskainen2003biased, ovaskainen2004habitat}.

The model of fragmented ecological systems benefits from a network-based approach - widely used in ecology and network science, with standard formulations found in \cite{newman_book, estrada_book, albert2002statistical, boccaletti2006complex} - where habitat patches are treated as nodes and dispersal pathways as edges, naturally capturing structural constraints and landscape connectivity \cite{urban2001landscape, minor2008graph, bascompte2007networks, ulanowicz2004quantitative}. Within this framework, random walks and diffusion processes on networks offer a powerful and flexible tool for studying movement-driven ecological phenomena \cite{masuda2017random, castellano_statistical_2009, pastor2001epidemic_1}. Notably, heterogeneous network structures can induce effective nonlinearities in otherwise linear dynamics due to finite-size effects such as crowding, as demonstrated in \cite{PhysRevLett.120.158301}. These emergent constraints were later generalized in \cite{carletti2020nonlinear}, where nonlinear walkers were shown to explore congested networks more efficiently, revealing rich dynamics driven purely by spatial limitations. These models capture essential features of dispersal in heterogeneous landscapes and can be extended to incorporate site-specific constraints and nonlinearities \cite{de2024multigraph, siebert2022nonlinear, de2022self, de2024emergence}.

Building on previous studies of nonlinear diffusion and constrained movement on networks, this work extends the framework to consider multiple competing species in a fragmented environment with limited resources. The model incorporates carrying capacity constraints at each patch and nonlinear dispersal behavior, where movement response depends on crowding. Following the mechanism introduced in~\cite{carletti2020nonlinear}, competition is modeled through an asymmetric perception of available resources: the more competitive species tends to overestimate local availability—adopting a more aggressive, optimistic strategy—while the less competitive species underestimates it, reacting more conservatively. This asymmetry shapes how each species responds to crowding and space occupation in the shared landscape. Early theoretical examples of diffusion constrained by local crowding include~\cite{fanelli2010diffusion}, where competition for space affects transport dynamics, and classical exclusion processes~\cite{liggett1999stochastic}, which model constrained stochastic interactions among individuals.

The analysis reveals that such competitive imbalance, combined with network topology, leads to spatial segregation between the species. The more competitive species preferentially occupies the central, highly connected patches, effectively displacing the weaker species to the periphery. As competition intensifies, the less competitive species abandons the network core and survives only in marginal areas with fewer connections. This emergent structure underscores the interplay between nonlinear movement, local constraints, and competitive dynamics, and offers a framework to explore how ecological outcomes such as exclusion or coexistence depend on both spatial structure and behavioral asymmetries.

\section{Dispersal of Space-Competing Species in Heterogeneous Environments}
\label{Sec:Model}

We begin by modeling the movement of individuals across a network of habitat patches, where each node \( i \) represents a spatial location with limited resources. Each patch is compartmentalized into \( N \) discrete slots, which define its carrying capacity: each slot can be occupied by either a single individual of species \( X \), a single individual of species \( Y \), or remain empty. We refer to the latter as \( E \), representing the slack compartment that accounts for available space. Movement is only possible if the destination patch has at least one available slot—that is, one or more \( E \)-type compartments. This discrete formulation reflects the idea that individuals occupy space exclusively, and sets the foundation for a mesoscopic description of dispersal driven by local interactions. The movement of individuals from node \( i \) to node \( j \) is described by the following reactions:
\begin{equation}
\begin{aligned}
X_i + \sigma_x E_j &\xrightarrow{\mathcal{D}_x} X_j + E_i, \\
Y_i + \sigma_y E_j &\xrightarrow{\mathcal{D}_y} Y_j + E_i,
\end{aligned}
\label{eq:reactions}
\end{equation}
where \( \mathcal{D}_x \) and \( \mathcal{D}_y \) are the diffusion coefficients, and \( \sigma_x \), \( \sigma_y \) are distortion parameters that reflect how strongly each species perceives the available space at the destination. When \( \sigma > 1 \), the species tends to overestimate the availability of resources, exhibiting an aggressive or optimistic movement strategy; conversely, \( \sigma < 1 \) reflects a cautious or conservative perception. These asymmetric responses underlie a form of indirect competition: individuals of each species attempt to occupy favorable locations based on their own perception of crowding, which influences movement success and ultimately drives spatial segregation. Thus, the parameters \( \sigma_x \) and \( \sigma_y \) encode behavioral traits that modulate how resource competition is expressed through space-limited migration.

We begin by analyzing the dynamics of species \( X \), focusing solely on its evolution while treating the presence of species \( Y \) as a given background. Starting from the individual-level interaction rules, we apply the law of mass action to derive a mesoscopic master equation that captures the average behavior of the system. Let \( nX_i(t) \) denote the number of individuals of species \( X \) present at node \( i \) at time \( t \), and let \( N \) be the number of discrete slots (i.e., the carrying capacity) at each node. We define the normalized density as
\(
x_i(t) = \lim_{N \to \infty} n X_i/N,
\)
and similarly for \( y_i(t) \). The term \( 1 - x_i - y_i \) then represents the average availability of space at node \( i \). Under these assumptions, the nonlinear diffusion equation governing the dynamics of \( x_i \) reads:
\begin{equation}  
\dot{x}_i = \mathcal{D}_x \sum_j A_{ij} \left[\frac{x_j}{k_j}\left(1 - x_i - y_i\right)^{\sigma_x} - \frac{x_i}{k_i}\left(1 - x_j - y_j\right)^{\sigma_x} \right],
\label{eq:onespecies}
\end{equation}
where \( A_{ij} \) denotes the entry of the adjacency matrix, equal to 1 if nodes \( i \) and \( j \) are connected and 0 otherwise, \( k_i = \sum_j A_{ij} \) is the degree of node \( i \) (i.e., the number of its neighbors), and \( \sigma_x \) is a scalar coefficient that quantifies the nonlinearity in species \( X \)’s response to perceived resource availability. This expression describes the net current of individuals entering and leaving patch \( i \), with each term accounting for dispersal between connected patches modulated by local density and limited space availability. It can be viewed as the deterministic approximation - or mesoscopic master equation - arising from stochastic dynamics constrained by finite space and non-linear competition for occupancy~\cite{gardiner}. 
To facilitate analytical treatment, we recast the system in a more compact operator form by introducing the random walk graph Laplacian, defined as \({L}^{\mathrm{RW}}_{ij} = {A_{ij}}/{k_j} - \delta_{ij} \)~\cite{newman_book}. This is achieved by adding and subtracting \( k_i \delta_{ij} \) to the adjacency matrix \( A_{ij} \), yielding the following reformulated dynamics:
\begin{equation}
\begin{aligned}
\dot{x}_i &= \mathcal{D}_x \sum_j {L}_{ij}^{\mathrm{RW}} \left[x_j \left(1 - x_i - y_i\right)^{\sigma_x} - \frac{k_j}{k_i} x_i \left(1 - x_j - y_j\right)^{\sigma_x} \right], \\
\dot{y}_i &= \mathcal{D}_y \sum_j {L}_{ij}^{\mathrm{RW}} \left[y_j \left(1 - x_i - y_i\right)^{\sigma_y} - \frac{k_j}{k_i} y_i \left(1 - x_j - y_j\right)^{\sigma_y} \right]
\end{aligned}
\label{eq:rwlaplacian_both}
\end{equation}
for all nodes \( i \), where, as will be shown below, \( \mathcal{D}_x \) and \( \mathcal{D}_y \) merely rescale the system’s temporal dynamics. This formulation emphasizes the diffusive structure of the dynamics and enables a systematic analytical exploration of spatial patterns and steady states on complex networks. For completeness, we have extended the formulation to include the dynamics of species \( Y \). 
When \( \sigma_x = \sigma_y = 1 \) and all degrees \( k_i \) are equal (i.e., the network is regular), Eqs.~\eqref{eq:rwlaplacian_both} reduces to a nonlinear diffusion model in which crowding effects arise purely from the presence of two competing species, as in~\cite{fanelli2010diffusion}. In our framework, this is generalized by incorporating distinct sources of nonlinearity: one due to interspecific competition for limited space, the other induced by topological heterogeneity in the network structure, as shown to affect diffusion in~\cite{PhysRevLett.120.158301}. At the opposite extreme, when \( \sigma_x = \sigma_y = 0 \), the dynamics reduce to standard linear random walk diffusion on networks~\cite{newman_book}. These limiting cases underscore how nonlinear interactions and network heterogeneity jointly enrich the system’s dynamics.

\begin{figure*}
    \centering
        \includegraphics[width=\textwidth]{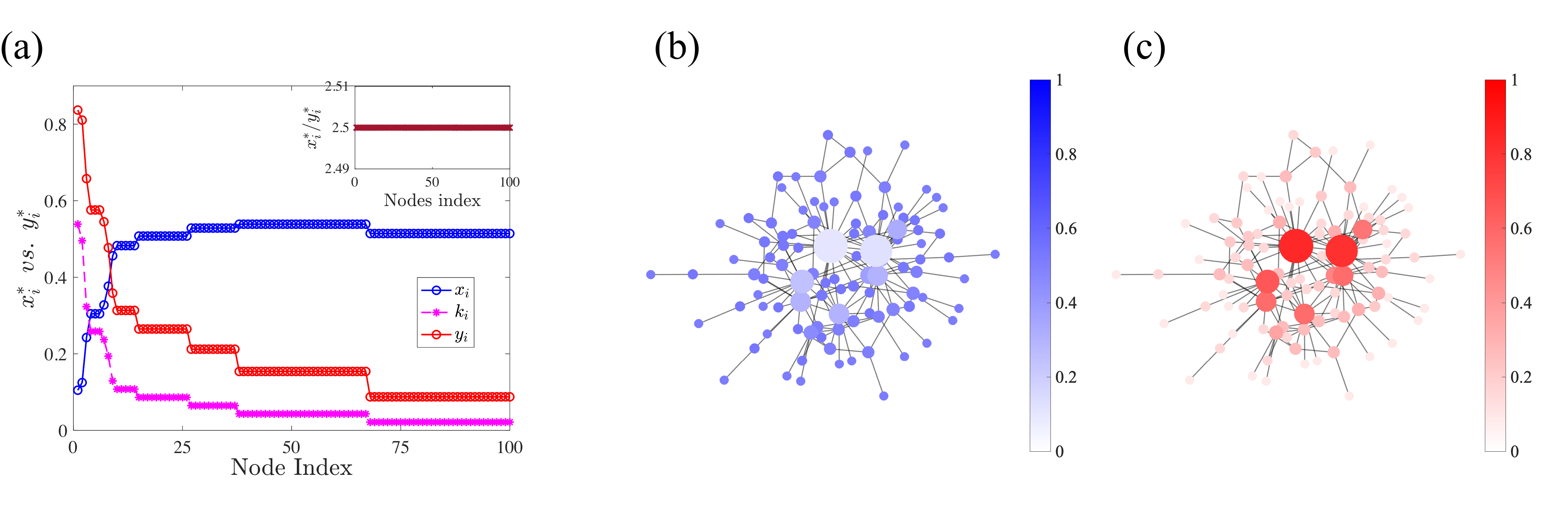} 
\caption{Representation of the species' final concentrations at steady state. 
(a) Final concentrations of \( x \) (blue) and \( y \) (red) across nodes, along with the normalized degree (magenta). 
The inset shows the ratio of the asymptotic densities \( x_i/y_i \) at each node for \( \sigma_x = \sigma_y = 1 \). 
(b) and (c) Network layouts showing the absolute concentrations of species \( x \) and \( y \), respectively. 
In both (b) and (c), node size is proportional to degree, with larger nodes representing more connected (central) patches. 
Darker colors indicate higher species concentration, while lighter colors indicate lower values. 
The network is a scale-free graph generated using the Barabási--Albert (BA) model with parameters \( N = 100 \), \( m_0 = 5 \), and \( m = 3 \), 
where \( N \) is the total number of nodes, \( m_0 \) the initial seed size, and \( m \) the number of preferentially attached links each new node forms. 
Parameters for the main panels: \( \sigma_x = 2.5 \), \( \sigma_y = 0.5 \), \( \mathcal{D}_x = \mathcal{D}_y = 30 \); 
initial densities are \( x = 0.5 \) and \( y = 0.2 \) uniformly across all nodes.}
    \label{fig:two_panel}
\end{figure*}

\section{Spatial Segregation Driven by Resource Competition}
\label{Sec:Segregation}

Having established the mechanistic model of competition and space-limited dispersal, we now investigate its behavior on a structured habitat network. As already mentioned, habitat fragmentation leads to spatial heterogeneity in connectivity among patches. To capture this structural complexity, we represent the habitat as a scale-free network generated via the Barabási–Albert (BA) model, where a small number of highly connected nodes (hubs) form the core of the network, while the periphery consists of sparsely connected patches~\cite{barabasi1999emergence}. Although all nodes share equal local carrying capacity, the more central nodes are considered ecologically advantageous due to their greater connectivity to surrounding patches—reflecting a higher likelihood of access to resources or mates, as commonly observed in fragmented habitats.

We examine the asymptotic densities of the two species by numerically integrating Eqs.~\eqref{eq:rwlaplacian_both} on a scale-free network, focusing on the effects of asymmetric competition. Specifically, when species \( X \) is less competitive than species \( Y \), that is, \( \sigma_x > \sigma_y \), individuals of species \( X \) are progressively displaced from the central, highly connected patches and relocate toward the periphery, while species \( Y \) occupies the core of the network. This outcome, shown in Fig.~\ref{fig:two_panel}(a), occurs despite species \( X \) (blue) being more abundant overall compared to \( Y \) (red), underscoring the strength of competitive asymmetry in spatial exclusion. Panels~\ref{fig:two_panel}(b) and \ref{fig:two_panel}(c) present the corresponding network representations for species \( X \) and \( Y \), respectively, where node size is proportional to degree, making the connectivity hierarchy visually explicit. In contrast, when both species have equal competitive ability (\( \sigma_x = \sigma_y = 1 \)), the inset of Fig.~\ref{fig:two_panel}(a) shows that the ratio \( x_i/y_i \) remains constant throughout the network, indicating that segregation is absent and the species densities depend only on the degree of the node. Reversing the asymmetry (\( \sigma_y > \sigma_x \)) leads to the opposite spatial arrangement. Together, this form of competitive sorting, illustrated in Fig.~\ref{fig:two_panel}, demonstrates how behavioral asymmetries in space-limited movement can drive spontaneous spatial segregation in heterogeneous environments.

\subsection{Derivation of Steady-State Solutions}

The equations introduced in~\eqref{eq:rwlaplacian_both} provide a suitable formalism to analyze the asymptotic solutions of the system. To generalize their structure, we rewrite the dynamics in terms of a node-specific interaction function \( f \), abstracting the bracketed terms. Denoting \( x_i^* \) and \( y_i^* \) as the steady-state densities of species \( X \) at node \( i \), Eqs.~\eqref{eq:rwlaplacian_both} can be expressed in the form:
\begin{equation}
    \dot{x}_i = \sum_j L_{ij}^{\mathrm{RW}} f(x_i,x_j,y_i,y_j),
    \label{generalxform}
\end{equation}
where for our system:
\begin{equation*}
    f(x_i,x_j,y_i,y_j) = x_j (1 - x_i - y_i)^{\sigma_x} - \frac{k_j}{k_i} x_i (1 - x_j - y_j)^{\sigma_x}.
    \label{specificxfform}
\end{equation*}
At steady state, \( \dot{x}_i = 0 \), so the right-hand side of Eq.~\eqref{generalxform} must vanish. Given that the graph is connected and \( k_j \neq 0 \) for all \( j \), the fixed-point condition requires:
\begin{equation}
    f(x_i^*,x_j^*,y_i^*,y_j^*) = a_i k_j,
    \label{fpeqn}
\end{equation}
where \( a_i \) is independent of \( j \). This follows from the steady-state condition imposed by the structure of the random walk Laplacian \( \boldsymbol{\mathit{L}}^{\mathrm{RW}} \)~\cite{newman_book}. In particular, evaluating this expression for \( i = j \) yields
\[
f(x_i^*,x_i^*,y_i^*,y_i^*) = a_i k_i = 0\,.
\]
Since \( k_i \neq 0 \), it follows that \( a_i = 0 \) for all \( i \). This confirms that the function \( f \) vanishes identically at steady state and ensures the internal consistency of the fixed-point condition.
The same derivation applies analogously to species \( Y \), yielding qualitatively equivalent expressions. Setting \( f = 0 \) in Eq.~\eqref{generalxform} and evaluating the fixed-point condition leads to the identity:
\begin{equation*}
\frac{x_i^*}{k_i (1 - x_i^* - y_i^*)^{\sigma_x}} = \frac{x_j^*}{k_j (1 - x_j^* - y_j^*)^{\sigma_x}} = C_x,
\label{xequal}
\end{equation*}
where \( C_x > 0 \) is a constant independent of node index. Due to the nonlinear dependence on \( x_i^* \) and \( y_i^* \), the fixed points must be determined implicitly. Rearranging
\begin{equation}
    \frac{x_i^*}{C_x k_i} = (1 - x_i^* - y_i^*)^{\sigma_x},
    \label{xequal2}
\end{equation}
and similarly for species \( Y \),
\begin{equation}
    \frac{y_i^*}{C_y k_i} = (1 - x_i^* - y_i^*)^{\sigma_y},
    \label{yequal}
\end{equation}
where \( C_y > 0 \) is the analogous constant for species \( Y \). These expressions define the steady-state densities implicitly and are {validated numerically} to capture the system’s behavior, with \(C_x\) and \(C_y\) {uniform across nodes} and \textit{time-invariant} after transients (long-time simulations; details omitted for brevity). The resulting fixed-point system is not analytically tractable in closed form, as it involves coupled nonlinear equations. Instead, it generalizes the structure proposed in~\cite{carletti2020nonlinear}, where similar implicit relationships arise from space-limited competition, here extended to incorporate asymmetric perception and network heterogeneity.

A particularly tractable case arises when \( \sigma_x = \sigma_y \), where the symmetry between species simplifies the fixed-point condition. Taking the right-hand sides of Eqs.~\eqref{xequal2} and~\eqref{yequal} under this assumption, we obtain:
\begin{equation}
\frac{x_i^*}{y_i^*} = \frac{C_x}{C_y}.
\label{eq:xyratio}
\end{equation}
This result implies that the ratio \( x_i^*/y_i^* \) is constant across the network—a prediction confirmed numerically and illustrated in the inset of Fig.~\ref{fig:two_panel}(a).

Building on the implicit fixed-point relations in Eqs.~(\ref{xequal2})–(\ref{yequal}) and the segregation mechanism discussed above, a sharp prediction emerges: {true node-level vacancies for species \(x\) occur only when \(\sigma_x>1\), and are amplified under the asymmetry \(\sigma_x>\sigma_y\)} (see Appendix~\ref{app:A} where this analytical criterion is borne out numerically in Fig.~\ref{fig:enter-label}: panel~(a), with sublinear responses \(\sigma_x,\sigma_y<1\), displays mild segregation without empty nodes, whereas panel~(b), with \(\sigma_x>1\) and \(\sigma_x>\sigma_y\), exhibits hub-level expulsion—species \(x\) becomes vanishingly rare on the highest-degree nodes—while its competitor consolidates in the core. The inset layouts, node size \(\propto\) degree, underscore that connectivity alone organizes the observed core–periphery split.).

\subsection{First-Order Expansion in the Sublinear Regime}

An explicit approximation for the steady-state densities \( x_i^* \) and \( y_i^* \) can be derived using a binomial expansion, which is valid for expressions of the form \( (1 - x)^\alpha \) when \( |x| < 1 \) and \( |\alpha x| \ll 1 \). This condition applies here since \( x_i^* + y_i^* < 1 \) by construction, due to the constraint of finite carrying capacity. Crucially, the approximation holds when both \( \sigma_x < 1 \) and \( \sigma_y < 1 \), ensuring that the nonlinear response to space remains close to linear. Ecologically, this regime corresponds to conservative species behavior, where the perceived availability of space is underestimated, leading to restrained dispersal toward crowded patches. In this setting, the binomial expansion
\(
(1 - x)^\alpha \approx 1 - \alpha x
\)
provides a tractable first-order approximation to model nonlinear movement under mild crowding effects.

Applying the approximation to the fixed-point equations yields
\begin{equation}
\frac{x_i^*}{C_x k_i} \approx 1 - \sigma_x \bigl(x_i^* + y_i^*\bigr), 
\qquad
\frac{y_i^*}{C_y k_i} \approx 1 - \sigma_y \bigl(x_i^* + y_i^*\bigr),
\label{eq:proxy_dense}
\end{equation}
and solving the resulting linear system gives
\begin{equation}
\begin{aligned}
x_i^* &= \frac{C_x k_i \bigl[ 1 + k_i C_y (\sigma_y - \sigma_x) \bigr]}{1 + k_i (\sigma_x C_x + \sigma_y C_y)}, \\[.4em]
y_i^* &= \frac{C_y k_i \bigl[ 1 + k_i C_x (\sigma_x - \sigma_y) \bigr]}{1 + k_i (\sigma_x C_x + \sigma_y C_y)}.
\end{aligned}
\label{eq:approx_fixed_points}
\end{equation}

Both densities share the same positive scaling factor,
\[
\frac{k_i}{1 + k_i (C_x \sigma_x + C_y \sigma_y)},
\]
so degree dependence is governed by the numerators. To expose the degree dependence in the numerators, introduce a sufficiently small asymmetry
\(\epsilon=\sigma_x-\sigma_y>0\) and define the rescaled quantities
\[
\hat{x}_i^* = C_x \bigl(1 - k_i C_y \epsilon\bigr), 
\qquad 
\hat{y}_i^* = C_y \bigl(1 + k_i C_x \epsilon\bigr).
\]
Then
\begin{equation}
\Delta(k_i) := \hat{y}_i^* - \hat{x}_i^* 
= (C_y - C_x) + 2 k_i C_x C_y \,\epsilon,
\label{deltaresult_new}
\end{equation}
which is strictly increasing in \(k_i\).

Equation~\eqref{deltaresult_new} implies a degree-driven inversion of dominance can occur. 
If initially $\Delta(0)=C_y-C_x<0$—i.e., $\hat{x}_i^*>\hat{y}_i^*$ as implied by \eqref{eq:proxy_dense}—then $\Delta(k_i)$ increases linearly with $k_i$ and crosses zero at
\begin{equation}
k_c=\frac{C_x-C_y}{2\,C_x C_y\,(\sigma_x-\sigma_y)},
\label{eq:crit}
\end{equation}
at which \(\Delta\) changes sign. For \(k_i<k_c\) one has \(\Delta<0\) (the \(X\)-density exceeds \(Y\) on low-degree nodes), whereas for \(k_i>k_c\) one has \(\Delta>0\) (the \(Y\)-density becomes larger on high-degree nodes), with the contrast increasing monotonically in \(k_i\) until the common scaling factor saturates.  By symmetry, reversing the sign of \(\epsilon\) exchanges the roles of \(X\) and \(Y\). In finite networks, this degree-controlled inversion is most prominent in hub-like or star-like topologies, while in the large-size limit the monotonicity and sign behavior follow directly from \(\Delta(k_i)\). This analysis is numerically validated in Fig.~\ref{fig:enter-label}(a) of the Appendix.

\section{A Generalized Reaction-Diffusion Framework}

So far, we have focused on the case of dispersal alone, examining how interspecific competition for limited resources drives the spatial segregation of two species across the network. This framework captured the essential role of mobility and crowding in shaping habitat partitioning but neglected local demographic dynamics and the potential for more complex, multi-species interactions. To render the model more ecologically realistic, we now extend the framework to include \( M \) interacting species and, crucially, incorporate a reaction term representing local population growth and decline. Specifically, we introduce {reaction--diffusion dynamics}, wherein individuals not only disperse through the network but also experience demographic changes at each node due to local interactions. In particular, the density \( x^{(\eta)}_i \) of species \( \eta \) at node \( i \) evolves according to
\begin{align}
\dot{x}^{(\eta)}_i 
&= r\,x^{(\eta)}_i\bigl(x^{(\eta)}_i - \mathcal{A}\bigr)
   \left(1-\sum_{\mu=1}^M x^{(\mu)}_i\right) +\nonumber\\
&\quad + \mathcal{D}_\eta \sum_{j} L^{\mathrm{RW}}_{ij} \Bigl[
   x^{(\eta)}_j x^{(\eta)}_i
   \left(1-\sum_{\mu=1}^M x^{(\mu)}_i\right)^{\sigma_\eta} +\nonumber\\
&\qquad\qquad
   - \frac{k_j}{k_i}\,x^{(\eta)}_i x^{(\eta)}_j
   \left(1-\sum_{\mu=1}^M x^{(\mu)}_j\right)^{\sigma_\eta} \Bigr],
\qquad \forall\, i,\eta. \label{eq:multi}
\end{align}
Here, the first term on the RHS describes the reaction dynamics: local logistic growth with an \textit{Allee effect} \cite{allee1927animal,allee1932studies,courchamp2008allee}, parameterized by $\mathcal{A}$. The Allee effect denotes {positive density dependence at low densities}: below the critical threshold $\mathcal{A}$ the per-capita growth rate becomes negative and the population deterministically declines (strong Allee effect), while above it growth is self-sustaining. Biologically, this captures mechanisms such as mate limitation, cooperative defense/foraging, and other forms of social facilitation. The second term captures the nonlinear {diffusion}, shaped by both crowding inhibition and a newly introduced term representing mutualistic bias: individuals preferentially move toward neighboring sites already occupied by conspecifics. The later as we will see in the following reinforces the segregation further. This behavior reflects ecologically relevant mechanisms such as gregarious movement, resource tracking, or habitat preference. As before, \( i \) indexes the habitat patch and \( \eta \in \{1, \dots, M\} \) denotes the species. The resulting dynamics ensure that the occupancy of each node evolves under the combined influence of competitive interactions, limited carrying capacity, and demographic thresholds, leading to segregation patterns that emerge from the interplay of reaction and diffusion processes.

\begin{figure*}
  \centering
  \makebox[\textwidth][l]{%
    \hspace{-1.2cm}%
    \includegraphics[width=1.1\textwidth]{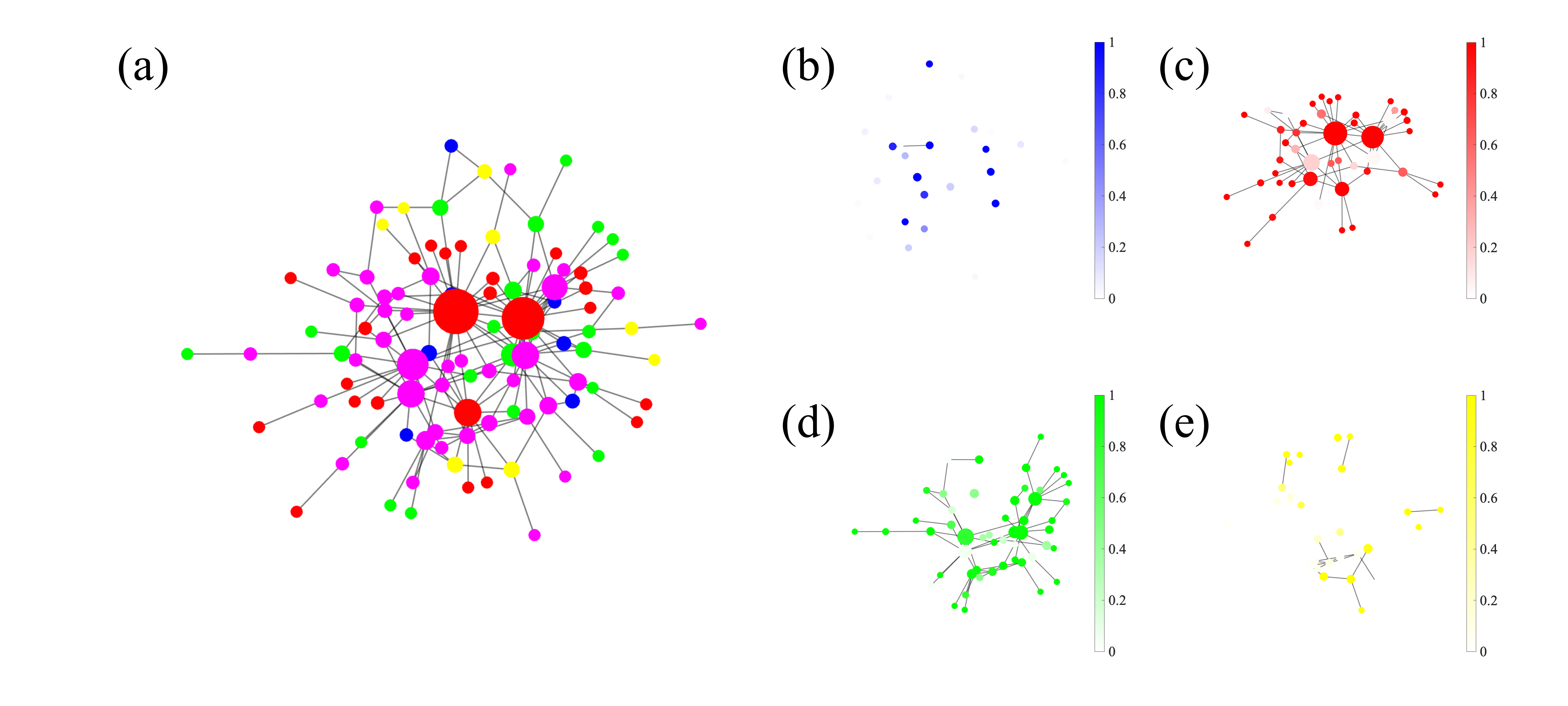}
  }
    \vspace{-1.cm}
\caption{Long-term dynamics of a four-species reaction–diffusion model on a network. Panel~(a) shows the global population distribution, with nodes colored by the dominant species: $x$ (blue), $y$ (red), $z$ (green), and $w$ (yellow). Nodes in magenta indicate coexistence of at least two species. Panels~(b)–(e) depict the subnetworks occupied primarily by species $x$, $y$, $z$, and $w$, respectively. All species are initialized at a density of $0.05$. Diffusion coefficients are set to $\mathcal{D}_x = 5000$, $\mathcal{D}_y = 100$, $\mathcal{D}_z = 500$, and $\mathcal{D}_w = 300$; crowding exponents are $\sigma_x = 4.5$, $\sigma_y = 1.1$, $\sigma_z = 1.3$, and $\sigma_w = 1.8$. The reaction rate is $r = 1$, and the Allee threshold is set to $\mathcal{A} = 0.15$ for all species, consistent with observational \emph{in situ} studies \citep{shepherd1995studies}.}
\label{fig:fourspecies}
\end{figure*}

Figure~\ref{fig:fourspecies} illustrates the emergent spatial organization in a system of four competing species evolving under the generalized reaction–diffusion dynamics. Panel~(a) shows the full network, with each node colored according to the dominant species: blue, red, green, or yellow. Nodes where two or more species coexist appear in magenta, highlighting areas of overlap. These segregation patterns arise from the combined effects of nonlinear diffusion and the {Allee effect}, further reinforced by mutualism, which favor settlement in regions already occupied by conspecifics. Panels~(b)–(e) display the subnetworks of nodes predominantly occupied by species $x$, $y$, $z$, and $w$, respectively.  

Stronger competitors—those characterized by reduced crowding sensitivity—occupy broad, contiguous domains, often concentrated around high-degree nodes. In contrast, weaker species are relegated to the network periphery, forming sparser and more fragmented subnetworks. The degree of fragmentation increases as competitiveness declines, as reflected by the proliferation of disconnected components and isolated patches in Fig.~\ref{fig:fourspecies}(b)-(e). The interplay between network topology and competitive hierarchy thus drives species-specific spatial distributions, producing structured segregation patterns with potential implications for long-term coexistence and evolutionary divergence.

We next develop a slow–fast analytical framework first determining the admissible configurations that explains the rich mosaic of diffusion–only fixed points. Once on the slow manifold, the species compete to determine node level dominance.

\subsection{Diffusion-only fixed points}

We start by analyzing steady states of the diffusion–only operator in \eqref{eq:multi}
\[
\dot{x}^{(\eta)}_i
= \mathcal{D}_\eta \sum_{j} L^{\mathrm{RW}}_{ij}
\Bigl[
x^{(\eta)}_j x^{(\eta)}_i\, s_i^{\sigma_\eta}
- \frac{k_j}{k_i}\,x^{(\eta)}_i x^{(\eta)}_j\, s_j^{\sigma_\eta}
\Bigr]=0,
\]
where \(s_i:=1-\sum_{\mu=1}^M x^{(\mu)}_i\).
If a species \(\eta\) is absent at node \(i\) (i.e., \(x_i^{(\eta)}=0\)), all fluxes involving \(i\) vanish identically for \(\eta\), so the diffusion term imposes no constraint there; in particular, a totally empty node (\(x_i^{(\mu)}=0\ \forall \mu\), hence \(s_i=1\)) is a diffusion fixed point.
On the other hand, when \(\eta\) is present at both ends of an edge \((i,j)\) and both nodes are unsaturated (\(x_i^{(\eta)}>0\), \(x_j^{(\eta)}>0\), \(s_i>0\), \(s_j>0\)), the net exchange must vanish, so the bracketed flux is zero and
\[
x^{(\eta)}_j x^{(\eta)}_i\, s_i^{\sigma_\eta}
-\frac{k_j}{k_i}\,x^{(\eta)}_i x^{(\eta)}_j\, s_j^{\sigma_\eta}=0
\;\Longleftrightarrow\;
k_i\,s_i^{\sigma_\eta}=k_j\,s_j^{\sigma_\eta}=: \tilde{C}_\eta .
\]
Thus, on each connected component of the {unsaturated support} of \(\eta\) (nodes where \(x^{(\eta)}>0\) and \(s>0\)) one has
\[
s_i=\Bigl(\frac{\tilde C_\eta}{k_i}\Bigr)^{1/\sigma_\eta}.
\]
If two species \(\eta,\zeta\) are both present at the same node \(i\) with \(s_i>0\), comparing the corresponding relations at two such nodes \(i\) and \(j\) yields
\[
\Bigl(\frac{1}{\sigma_\eta}-\frac{1}{\sigma_\zeta}\Bigr)\,(\log k_j-\log k_i)=0.
\]
Hence, across any edge $(i,j)$ with $s_i,s_j>0$, \emph{unsaturated coexistence} of two species at both endpoints is possible only if either $\sigma_\eta=\sigma_\zeta$ or $k_i=k_j$; otherwise, at least one endpoint must lose one species or become saturated. However, this does not exclude cases where one endpoint is mixed and the other is single-species, since the flux of the absent species vanishes and imposes no constraint. This clearly illustrates how the mutualistic term contributes in reinforcing the segregation of the species accross the network. By contrast, \emph{fully packed} nodes ($s_i=0$) may host arbitrary mixtures, as all edge fluxes cancel irrespective of composition.

These constraints give rise to a spatial \emph{mosaic} of equilibria. Empty nodes (\(s_i=1\)) remain fixed points and act as passive boundaries. Unsaturated single-species nodes (\(s_i>0\)) form extended domains where \(s_i\propto k_i^{-1/\sigma_\eta}\), so that a species occupies preferentially the higher-degree nodes within its territory. Mixed nodes can appear either when saturation occurs (\(s_i=0\)) or when coexistence does not propagate beyond the node—typically when one species is present only locally, or when its neighbors are empty or host a different single species. The resulting steady state is therefore a patchwork of single-species unsaturated regions, locally mixed (possibly unsaturated) boundary nodes, and empty sites.

\begin{figure*}
  \centering
  \makebox[\textwidth][l]{%
    \hspace{-1.2cm}%
    \includegraphics[width=1.1\textwidth]{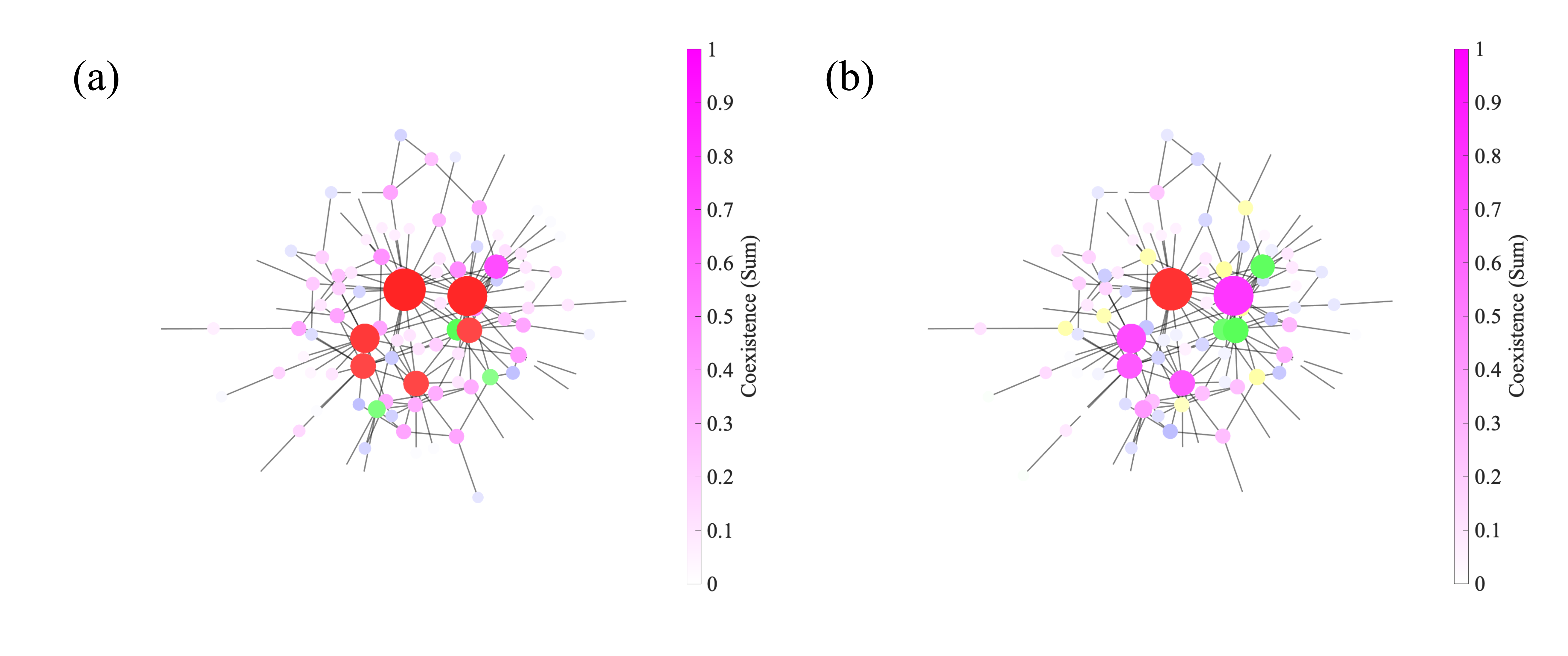}
  }
  \vspace{-1.cm}
\caption{Diffusion-only dynamics on the network and parameter set of Fig.~\ref{fig:fourspecies}. Nodes are colored by the dominant species: \(X\) (blue), \(Y\) (red), \(Z\) (green), \(W\) (yellow); magenta marks coexistence. (a) All species initialized at \(0.05\). (b) Identical setup except species \(Y\) starts at \(0.01\). Reaction terms are switched off (\(r=0\)). Even under pure diffusion, when all species start equally, the strongest competitor \(Y\) colonizes most hubs; with \(Y\) reduced, it fails to conquer the largest hub and many medium-degree nodes remain mixed, while the second-strongest \(Z\) invades several secondary hubs, in some cases displacing \(Y\). Note that the peripheral nodes shown in white (not visible at this color scale) are empty.}
\label{fig:fourspecies_diff}
\end{figure*}

Moreover, degree and the exponents \(\{\sigma_\eta\}\) bias {who} wins on unsaturated nodes. The steady–flux constraint
\(
s_i=(\tilde C_\eta)^{1/\sigma_\eta}\,k_i^{-1/\sigma_\eta}
\)
shows that, for each species \(\eta\), the compatible vacancy \(s_i\) {decreases monotonically with degree}. Thus a species concentrates preferentially on larger–degree nodes in proportion to the decay \(k_i^{-1/\sigma_\eta}\). Comparing two species \(\eta,\zeta\),
\[
\frac{s_i^{(\eta)}}{s_i^{(\zeta)}}=
\left(\frac{\tilde C_\zeta^{1/\sigma_\zeta}}{\tilde C_\eta^{1/\sigma_\eta}}\right)\,
k_i^{\,1/\sigma_\zeta-1/\sigma_\eta},
\]
so, for comparable \(\tilde C\)’s, the species with smaller \(\sigma\) has the steeper decay and is increasingly favored as \(k_i\) grows; the conserved–mass constants \(\tilde C_\eta\) shift, but do not remove, the generic \(k^{-1/\sigma}\) bias. When \(\sigma_\eta=\sigma_\zeta\), degree does not discriminate between the two.

Figure~\ref{fig:fourspecies_diff} illustrates how diffusion alone can generate heterogeneous spatial dominance in the absence of reaction terms. When all species are initialized at equal densities panel~(a), the most competitive species \(Y\) (red) rapidly colonizes the network hubs, establishing dominance across most high-degree nodes. In contrast, when \(Y\) starts from a smaller initial density than the other species panel~(b), it still occupies the largest hub, but fails to invade the second largest hubs which together with many medium-degree nodes remain in mixed states. Moreover, the second-strongest competitor \(Z\) (green) expands its range, occupying several secondary hubs and in some cases displacing \(Y\). Notably, the third-strongest species \(W\) (yellow) becomes more prevalent at intermediate-degree nodes that were previously impeded by \(Y\) in panel~(a), underscoring the sensitivity of the diffusion-driven steady state to small perturbations of the initial condition. Finally, peripheral nodes rendered in white (not visible at this scale) are vacant—i.e., no species are present—consistent with the theoretical prediction for the diffusion-only regime.

\subsection{Asymptotic structure and slow--fast selection of patterns}

On the fast timescale, i.e.\ $\mathcal{D}_\eta \gg r$ for all $\eta$, diffusion drives the system to a \emph{mosaic} constrained by the edgewise steady--flux relations derived above: unsaturated nodes ($s_i>0$) end up singly occupied (exactly one species has $x_i^{(\cdot)}>0$), fully packed nodes ($s_i=0$) can carry arbitrary mixtures, and some nodes may remain empty ($s_i=1$). Once this mosaic is formed, diffusion is silent on the slow timescale and each node evolves independently under the reaction term.

Consider a node $i$. For any present species $\mu$, the slow dynamics near the mosaic satisfy
\[
\operatorname{sign}\,\dot x_i^{(\mu)}=\operatorname{sign}\!\big(x_i^{(\mu)} - \mathcal{A}\big),
\]
so that
\begin{align*}
0<x_i^{(\mu)}<\mathcal{A} \ \Rightarrow\ &\dot x_i^{(\mu)}<0,\qquad 
x_i^{(\mu)}>\mathcal{A} \ \Rightarrow\ \dot x_i^{(\mu)}>0,\\[.5em]
x_i^{(\mu)}&\in\{0,\mathcal{A}\} \ \Rightarrow\ \dot x_i^{(\mu)}=0.
\end{align*}
Thus $x_i^{(\mu)}=0$ is absorbing and $x_i^{(\mu)}=\mathcal{A}$ is a threshold. Because a saturated node enforces $\sum_{\mu=1}^M x_i^{(\mu)}=1$, any attracting slow equilibrium at that node can contain only components with $x_i^{(\mu)}\ge \mathcal{A}$. If $m$ species persist, then
\[
1=\sum_{\text{survivors}} x_i^{(\mu)} \ \ge\ m\mathcal{A} \quad\Longrightarrow\quad m\le \lfloor 1/\mathcal{A}\rfloor.
\]
In generic outcomes, exactly $\lfloor 1/\mathcal{A}\rfloor$ species survive at level $x_i^{(\mu)}=\mathcal{A}$, with all other components extinct, where $\lfloor\cdot\rfloor$ denotes the greatest integer less than or equal to its argument. Any residual mass smaller than $\mathcal{A}$ cannot sustain an additional component, it would lie below threshold and decay, and is transiently redistributed before the node settles into this threshold--coexistence configuration.

If instead the node is singly occupied ($s_i>0$ and exactly one $x_i^{(\mu)}>0$), the slow dynamics are one–dimensional: starting below $\mathcal{A}$ leads to extinction at that node; starting above $\mathcal{A}$ drives the occupant to fill the node, eliminating vacancy and producing a pure saturated state. Similarly, empty nodes ($s_i=1$, all $x_i^{(\mu)}=0$) remain empty under the slow flow in the absence of external inputs.

In summary, fast diffusion first creates the mosaic—unsaturated single–species nodes, mixed nodes, and possibly empty nodes—while the slow reaction step filters it locally: mixed nodes retain at most $\lfloor 1/\mathcal{A}\rfloor$ components pinned at $\mathcal{A}$; unsaturated single–species nodes polarize to extinction or purity depending on their initial level relative to $A$; empty nodes stay empty.

\section{Conclusions and discussion}

Habitat fragmentation profoundly influences biodiversity by constraining dispersal and intensifying competition for limited resources. In such fragmented landscapes, species must navigate complex trade-offs between colonizing high-quality, densely connected patches and avoiding competitors, often resulting in spatial segregation and the formation of distinct habitat domains.  

In this work, we developed a degree-normalized, nonlinear dispersion-driven model on a network to investigate these segregation dynamics in fragmented habitats. The model incorporates local carrying capacities at each node---captured via crowding-induced nonlinearities quantified by distortion exponents $\sigma_\eta$---and population dynamics governed by an Allee effect. Analytical progress was first made in the two-species diffusion-only scenario, where the fixed points were characterized implicitly and approximated explicitly using a binomial expansion. This analysis identified the parameter regimes under which nodes are vacated, reflecting habitat fragmentation and spatial segregation. In particular, when both \(\sigma_x, \sigma_y < 1\), the two species tend to coexist within each node, with moderate segregation between peripheral and hub nodes. Conversely, complete exclusion of one species from a node requires the competitor to exhibit \(\sigma > 1\), amplifying competitive asymmetries.  

The framework was further extended to an arbitrary number of species (\(M > 2\)) and augmented with a reaction term modeling local growth and decline, incorporating demographic thresholds via the Allee effect. To further reinforce the species segregation, a mutualistic term was added to the diffusion dynamics. For this generalized reaction--diffusion model, the existence of stable, segregated spatial patterns was established analytically, confirming that such patterns persist beyond the two-species setting. Numerical simulations complemented the analysis, demonstrating how nonlinear diffusion, demographic thresholds, and network topology together structure species into distinct domains: stronger competitors concentrate on high-degree nodes, while weaker species occupy fragmented, peripheral subnetworks. 

These findings underscore the fundamental role of crowding effects, competitive asymmetries, and habitat structure in shaping biodiversity patterns in complex landscapes. In particular, habitat fragmentation—via reduced patch size, increased isolation, altered connectivity, and stronger edge effects—modulates dispersal, intensifies competitive filters, and drives scale-dependent diversity and stability.

\begin{acknowledgments}
M.A. acknowledges support from the SEED grant from the FSU Council on Research and Creativity SEED grant \textit{Structure and dynamics of nonnormal networks}.
\end{acknowledgments}

\appendix

\begin{figure*}
    \centering
    \includegraphics[width=\linewidth]{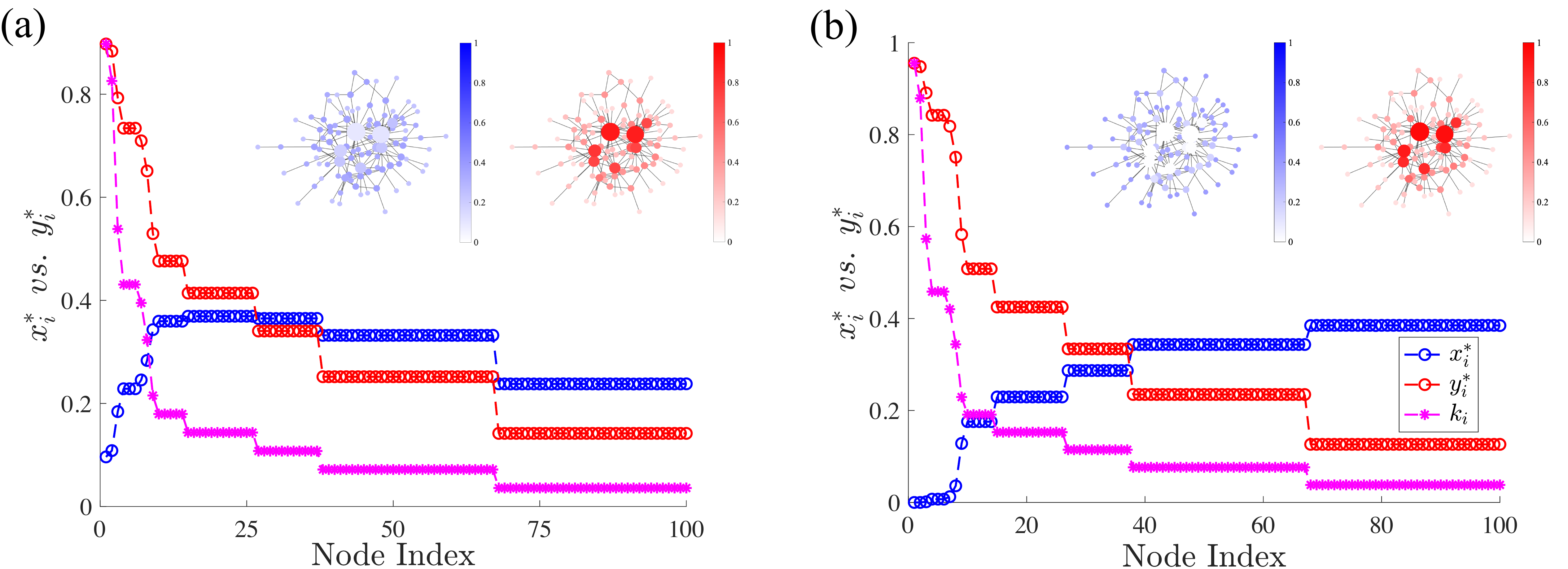}
    \caption{Representation of steady-state species distributions across the network. (a) Final concentrations of species \( x \) and \( y \) under sublinear movement responses, with \( \sigma_x = 0.9 \), \( \sigma_y = 0.3 \), and equal diffusion coefficients \( \mathcal{D}_x = \mathcal{D}_y = 30 \); initial densities are uniformly set to 0.3. (b) Same network structure, but with parameters chosen to emphasize strong asymmetry: \( \sigma_x = 5.5 \), \( \sigma_y = 0.5 \), and a much larger diffusion coefficient for species \( x \) (\( \mathcal{D}_x = 2000 \)) to compensate for its high perceived availability. In both panels, the inset shows the corresponding network layout, with node sizes proportional to degree. In panel (b), species \( x \) is nearly absent from high-degree nodes—such as nodes 1 and 2—where concentrations fall below \( 10^{-4} \).}
    \label{fig:enter-label}
\end{figure*}

\section{Necessary conditions for vacant nodes}
\label{app:A}

We now investigate the regimes under which vacant nodes arise; that is, the necessary conditions for a species to completely vacate a node.  

\vspace{1em}

{
Starting from the implicit fixed-point formulation, suppose there exists at least one node \( i \) such that \( x_i^* = 0 \) and \( y_i^* \neq 0,1 \).  
This assumption immediately implies
\[
C_x = \frac{x_i^*}{k_i (1-x_i^*-y_i^*)^{\sigma_x}} = 0,
\]
which forces \( C_x = 0 \) for all \( i \), and therefore \( x_i^* = 0 \) for all \( i \).  
This contradicts the initial conditions and implies that no mass of species \( x \) exists throughout the network, which is not possible.
}

{
Next, consider the case where there exists a node \( i \) such that \( x_i^* = 0 \) and \( y_i^* = 1 \).  
We first take the limit \( y_i^* \to 1^- \), which implies \( 1-x_i^*-y_i^* \to 0^+ \), since \(x_i^* \geq 0\) and \(y_i^* \leq 1\) at all times.  
Then we take \(x_i^* \to 0^+\), and in this regime \(1-x_i^*-y_i^*\) behaves like \(x_i^*\), remaining positive.  
To avoid ambiguity when raising to the power \(\sigma_x\), we initially write \(1-x_i^*-y_i^* = |1-x_i^*-y_i^*|\), although in this limit the absolute value equals simply \(1-x_i^*-y_i^*\).  
We write the limits explicitly as nested to make clear the order in which they are taken:
\begin{align}
    C_x &=
    \lim_{y_i^* \to 1^-} 
    \lim_{x_i^* \to 0^+} 
    \frac{x_i^*}{k_i |1-x_i^*-y_i^*|^{\sigma_x}} \nonumber\\
    &=
    \lim_{y_i^* \to 1^-} 
    \lim_{x_i^* \to 0^+} 
    \frac{x_i^*}{k_i (1-x_i^*-y_i^*)^{\sigma_x}}
    =
    \lim_{x_i^* \to 0^+} 
    \frac{1}{k_i (x_i^*)^{\sigma_x-1}}.
\end{align}
}

{
If \( \sigma_x > 1 \), then \( x_i^*{}^{\sigma_x-1} \to 0 \) as \( x_i^* \to 0^+ \), so \( C_x \to \infty \) unless \( k_i \to \infty \) fast enough to compensate.  
If \( \sigma_x < 1 \), then \( x_i^*{}^{\sigma_x-1} \to \infty \), so \( C_x \to 0 \).  
If \( \sigma_x = 1 \), then \( C_x = 1/k_i \), which is finite.
}
We therefore consider the joint limit
\[
C_x =
\lim_{\substack{x_i^* \to 0^+ \\ y_i^* \to 1^- \\ k_i \to \infty}} 
\frac{1}{k_i x_i^*{}^{\sigma_x-1}}.
\]
For \( \sigma_x < 1 \), this limit tends to zero.  
For \( \sigma_x > 1 \), it may converge if \( k_i \) grows appropriately to balance the vanishing \( x_i^* \), suggesting that \( \sigma_x > 1 \) is a necessary condition for empty nodes to occur.
Similarly, for \( C_y \), we find:
\[
C_y =
\lim_{\substack{x_i^* \to 0^+ \\ y_i^* \to 1^- \\ k_i \to \infty}} 
\frac{y_i^*}{k_i (1-x_i^*-y_i^*)^{\sigma_y}}.
\]
This limit may converge when \( 1 < \sigma_x > \sigma_y \), highlighting again the critical role of the exponents in determining the possibility of node vacancy. The condition \( \sigma_y < \sigma_x \) is assumed to reflect the weaker competitive ability of \( x \), highlighting again the critical role of the exponents in determining node vacancy.

Our analysis predicts that true node-level vacancies for species \(x\) arise only when \(\sigma_x>1\) and are further amplified under asymmetry \(\sigma_x>\sigma_y\). 
This prediction is numerically validated in Fig.~\ref{fig:enter-label}: panel~(a), with sublinear responses (\(\sigma_x,\sigma_y<1\)), shows mild segregation without empty nodes, whereas panel~(b), with \(\sigma_x>1\) and \(\sigma_x>\sigma_y\), exhibits hub-level expulsion—species \(x\) becomes vanishingly rare on the highest-degree nodes—while its competitor consolidates in the core. 
The inset layouts (node size \(\propto\) degree) underscore that connectivity alone organizes the observed core–periphery split.


\bibliographystyle{apsrev4-2}
\bibliography{NLRW_2species}

\end{document}